\documentclass[epj,nopacs]{svjour}
\usepackage{graphics}
\begin{document}
%
%\title{Exploring the nuclear matter phase diagram with heavy ion reactions}
\title{Probing the hadron-quark mixed phase at high isospin and baryon density}
 \subtitle{Sensitive observables}
 \dedication{ We like to dedicate
this work to the memory of Prof. Liu Bo, of the Institute for High
Energy Physics (IHEP) of Beijing, suddenly passed away few months
ago, for his continuous interest in the project and for the
important contributions. }
\author{Massimo Di Toro\inst{1,2}, Maria Colonna\inst{1}, Vincenzo Greco\inst{1,2}
 \and Guo-Yun Shao\inst{3}}
%\thanks{\emph{e-mail: ditoro@lns.infn.it}%
%}                     % Do not remove
\institute{Laboratori Nazionali del Sud, INFN, via Santa Sofia 62,
I-95123, Catania, Italy  \and Physics-Astronomy Dept., University of
Catania, Italy \and Physics Dept., Xi'an Jiaotong University, China}
\date{Received: date / Revised version: date}
\abstract{ We discuss the isospin effect on the possible phase
transition from hadronic to quark matter at high baryon density and
finite temperatures. The two-Equation of State (Two-EoS) model is
 adopted to describe the hadron-quark phase transition in
dense matter formed in heavy-ion collisions. For the hadron sector
we use Relativistic Mean Field (RMF) effective models, already
tested on heavy ion collision (HIC). For the quark phase we consider
various effective models, the MIT-Bag static picture, the
Nambu--Jona-Lasinio (NJL) approach with chiral dynamics and
finally the NJL coupled to the Polyakov-loop field (PNJL), which
includes both chiral and (de)confinement dynamics. The idea is to
extract mixed phase properties which appear robust with respect to
the model differences. In particular we focus on the phase transitions of
isospin asymmetric matter, with two main results: i)
an earlier transition to a mixed hadron-quark phase, at lower baryon
density/chemical potential with respect to symmetric matter; 
ii) an "Isospin Distillation" to the
quark component of the mixed phase, with predicted effects on the
final hadron production. Possible observation signals are suggested
to probe in heavy-ion collision experiments at intermediate
energies, in the range of the NICA program. }
%end of abstract
\mail{ditoro@lns.infn.it}
\titlerunning{Probing the hadron-quark mixed phase}

\maketitle

\section{Motivations}

In heavy ion collisions at intermediate beam energies in the AGeV
range, rather high density regions can be reached, opening the
possibility for new degrees of freedom to come into play. This kind
of collisions is usually described for hadronic matter 
 within relativistic mean-field(RMF) models and transport theories
 \cite{baranPR}. Here we want to show that
 in neutron-rich systems the transition from the nuclear
(hadron) to the quark deconfined
 (quark-gluon plasma) phase could take place even at density and
 temperature conditions reached along collision dynamics in the
 intermediate energy range.
Such transition in very isospin asymmetric matter is also of large interest
in the study of neutron stars (in the following we in short talk of  "asymmetric matter").

Hadronic matter is expected to undergo a phase
transition to a deconfined phase of quarks and gluons at large
densities and/or high temperatures. On very general grounds, the
critical densities of the transition should be dependent on the isospin
of the system, but no experimental tests of this dependence have
been performed so far. Moreover, up to now, data on the phase
transition have been extracted from ultrarelativistic collisions,
when large temperatures but low baryon densities are reached.
\begin{figure}
%\begin{center}
\resizebox{0.5\textwidth}{!}{%
  \includegraphics{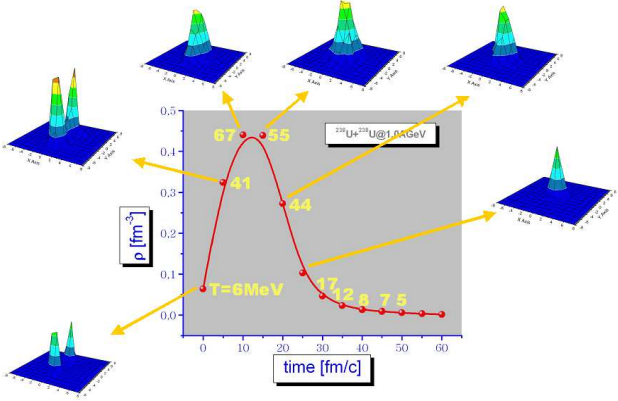}
 }
%\includegraphics[scale=0.36]{HQNICA1.ps}
%\vskip -1.3cm
%\includegraphics[angle=-90,scale=0.30,trim=0 0 -20 0]{UU_cm.ps}
\caption{\label{figUU} (Color on line) Uranium-Uranium $1~AGeV$
semicentral collision. Correlation between density, temperature,
momentum thermalization inside a cubic cell 2.5 $fm$ wide, located
in the center of mass of the system.}
 %%%
 %ts} - baryon density in $\rho_0$ units; {\it grey dots} -
%quadrupole moment in momentum space; {\it squares} - resonance
%density. } \vskip -1.7cm
%\end{center}
\end{figure}

In order to check the possibility of observing some precursor
signals of  new physics even in collisions of stable nuclei at
intermediate energies we have performed event simulations for
the collision of very heavy, neutron-rich elements. We have chosen
the reaction $^{238}U+^{238}U$ (average proton fraction $Z/A=0.39$)
at $1~AGeV$ and semicentral impact parameter $b=7~fm$ just to
increase the neutron excess in the interacting region
\cite{Toro06,Toro09}.
 To
evaluate the degree of local equilibration and the corresponding
temperature we have followed the momentum distribution in a space
cell located in the c.m. of the system; in the same cell we report
the maximum density evolution. Results are shown in
Fig.~\ref{figUU}. We see that after about $10~fm/c$ a nice local
equilibration is achieved.  We have a unique Fermi distribution and
from a simple fit we can evaluate the local temperature.
  At this beam energy the maximum density
 coincides with the thermalization, then the system is quickly cooling
 while expanding.
 We can extract the time evolution of all physics
 parameters inside the c.m. cell in the interaction region.
We find that a rather exotic nuclear matter is formed in a transient
time of the order of $10~fm/c$, with baryon density around $3-4$
times saturation density $\rho_0$, temperature $50-60~MeV$, energy
density $500~MeV~fm^{-3}$ and proton fraction between $0.35$ and
$0.40$,
 \cite{Toro06,Toro09}. This could be well inside the estimated mixed
phase region, as discussed in the following.

\subsection{Isospin effect on the Hadron-Quark transition: simple arguments}

At high temperature $T$ and small quark chemical potential $\mu_q$
lattice-QCD (l-QCD) calculations provide a valuable tool to investigate
the transition to the hadronic phase.
%at high $T$ and small
%quark chemical potential $\mu_q$.
% ~\cite{Karsch01, Karsch02,
%Allton02, Kaczmarek05, Cheng06, YAoki09, Borsanyi10}.
The transition
appears of continuous type (crossover) with a critical temperature
$T_c$ around 170-180 MeV. Isospin and other properties of the hadron
interaction appear not relevant here ~\cite{Liu11}. The reason is that 
at low chemical 
potential only thermal excitations contribute to the pressure, 
ruled essentially by the particle degrees of freedom ~\cite{Yagi2005}. 

However the lattice calculations suffer serious problems
at large chemical potentials and the validity of the results at
$\mu_q/T_c > 1$ is largely uncertain ~\cite{Fukushima11}. Some
phenomenological effective models have been introduced, like the
MIT-Bag ~\cite{Chodos74} and the more sophisticated Nambu-Jona
Lasinio (NJL)~\cite{Nambu61,Buballa05} and Polyakov-NJL
(PNJL)~\cite{Fukushima04,Ratti06,Schaefer10} models, where the
chiral and deconfinement dynamics is accounted for effectively. We remark here
that only scalar interactions are generally considered in the quark
sector. In the PNJL case the transition at low-$\mu$ is well in
agreement with l-QCD results~\cite{Borsanyi14}, however
still important interaction channels of the hadron sector are not included and so the
expected transition at high baryon and isospin density cannot be
fully trusted.

In order to overcome the problem and to get some predictions about
the effect of the transition in compact stars
~\cite{Glendenning92,Glendenning98,Burgio02,Shao10,Shao110,Dexheimer101}
and high energy heavy ion collisions
~\cite{Toro06,Toro09,Liu11,Muller97,Torohq11,Cavagnoli10,Shao111,Shao112},
recently Two-EoS (Two-Equation of State) models have been
introduced where both hadron and quark degrees of freedom are
considered, with particular attention to the transition in 
asymmetric matter.

In this report the discussion is limited to nucleonic hadron matter and $u,d$ quark matter,
also consistently with the analysis of intermediate energy collisions, but including 
the possibility of an extension to 
strange particles at higher excitation energies ~\cite{Shao112}.

In isospin asymmetric fermionic systems we have a repulsive symmetry term coming from 
two contributions: kinetic, from the Fermi motion and potential, from the particle interaction
(isovector terms). In the nucleonic phase we have both contributions, with the potential one 
rather important at high densities ~\cite{hadint}. At variance, in the quark phase, in all the effective
QCD models, we have only the kinetic part, which is rather slowly increasing with the baryon density, being proportional to the Fermi energy ~\cite{baranPR}. Within this picture we can predict rather relevant 
isospin effects on the transition in asymmetric matter, which should be seen at the NICA energies:

i) Onset of the transition at lower densities due to the larger symmetry pressure in the hadron phase;

ii) Isospin enrichment of the quark component in the mixed phase due to the smaller symmetry repulsion
(Isospin Distillation).

Related observables are discussed in final section. Of course the measurements of the NICA project 
will be important also to shed lights on the relevance of isovector contributions in the quark sector.

\section{Mixed phase and isospin distillation}

When a mixed (coexistence) phase of quarks and hadrons is considered,
the Gibbs conditions (thermal, chemical and mechanical equilibrium)
\begin{eqnarray}\label{tcm}
& &\mu_B^H(\rho_B^{},\rho_3^{},T)=\mu_B^Q(\rho_B^{},\rho_3^{},T)\nonumber\\
& &\mu_3^H(\rho_B^{},\rho_3^{},T)=\mu_3^Q(\rho_B^{},\rho_3^{},T)\nonumber\\
& &P^H(\rho_B^{},\rho_3^{},T)=P^Q(\rho_B^{},\rho_3^{},T),
\end{eqnarray}
should be fulfilled ~\cite{Glendenning92}. In Eqs.~(\ref{tcm}),
$\rho_B^{}=(1-\chi)\rho_B^{H}+\chi \rho_B^{Q}$ is the mean baryon
density and  $\rho_3^{}=(\rho_p-\rho_n)=(1-\chi)\rho_3^{H}+\chi
\rho_3^{Q}$ is the isospin density, where $\chi$ is the quark
fraction. $\rho_B^{H,Q}$ and $\rho_3^{H,Q}$ ($\mu_B^{H,Q}$ and $\mu_3^{H,Q}$) are
baryon and isospin densities (chemical potentials), respectively, in
the two phases. $P^{H,Q}$ indicates the pressure in the two phases.

The asymmetry parameters for hadronic and quark matter are defined,
respectively,

\noindent
\begin{eqnarray}\label{eq:43}
\alpha^{H}\equiv-\frac{\rho_{3}^{H}}{\rho_{B}^{H}}=\frac{\rho_n - \rho_p}{\rho_n+\rho_p}~,~~~~
\alpha^{Q}\equiv-\frac{\rho_{3}^{Q}}{\rho_{B}^{Q}}=3\frac{\rho_d - \rho_u}{\rho_d+\rho_u},
\end{eqnarray}

The factor 3 in the quark case  comes from the corresponding baryon and isospin densities 
$\rho_{B}^{Q}=\frac{\rho_d+\rho_u}{3}$, $\rho_{3}^{Q}={\rho_u-\rho_d}$. For pure neutron matter
$\rho_d=2\rho_u$ and $\alpha^{Q}$ is consistently 1.

In heavy-ion collisions, for a given isospin asymmetry of the considered
experiment, the global asymmetry parameter $\alpha$
%in the coexistence phase
%
\begin{eqnarray}
& &   \alpha\equiv-\frac{\rho_{3}^{}}{\rho_{B}^{}}= - \frac{(1-\chi)\rho_3^{H}+
\chi \rho_3^{Q}}{(1-\chi)\rho_B^{H}+\chi \rho_B^{Q}},
\end{eqnarray}
remains constant  according to the charge conservation in the strong interaction,
but the local asymmetry parameters $\alpha^H,\alpha^Q$ 
in the
separate phases can vary with $\chi$, as determined by the energetically stable state of the system.
For details, one can
refer to Refs.~\cite{Torohq11,Shao111,Shao112}.

\begin{figure}[htbp]
%\begin{center}
\resizebox{0.4\textwidth}{!}{%
  \includegraphics{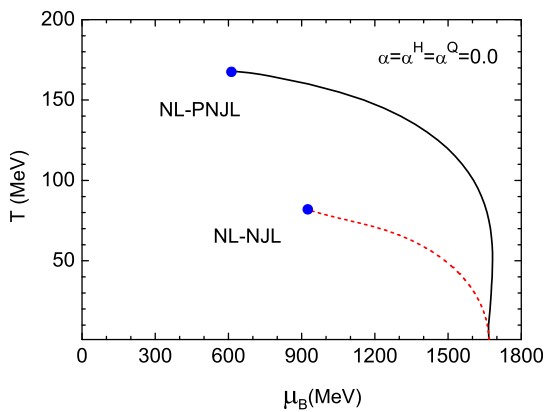}%{val1.ps}
}
\caption{\label{fig:T-Mu-with-delta-alpha=0} (Color on line) Phase
diagram in $T-\mu_B^{}$ plane for symmetric matter \cite{Shao112}.}
%\end{center}
\end{figure}
\begin{figure}[htbp]
%\begin{center}
\resizebox{0.4\textwidth}{!}{%
  \includegraphics{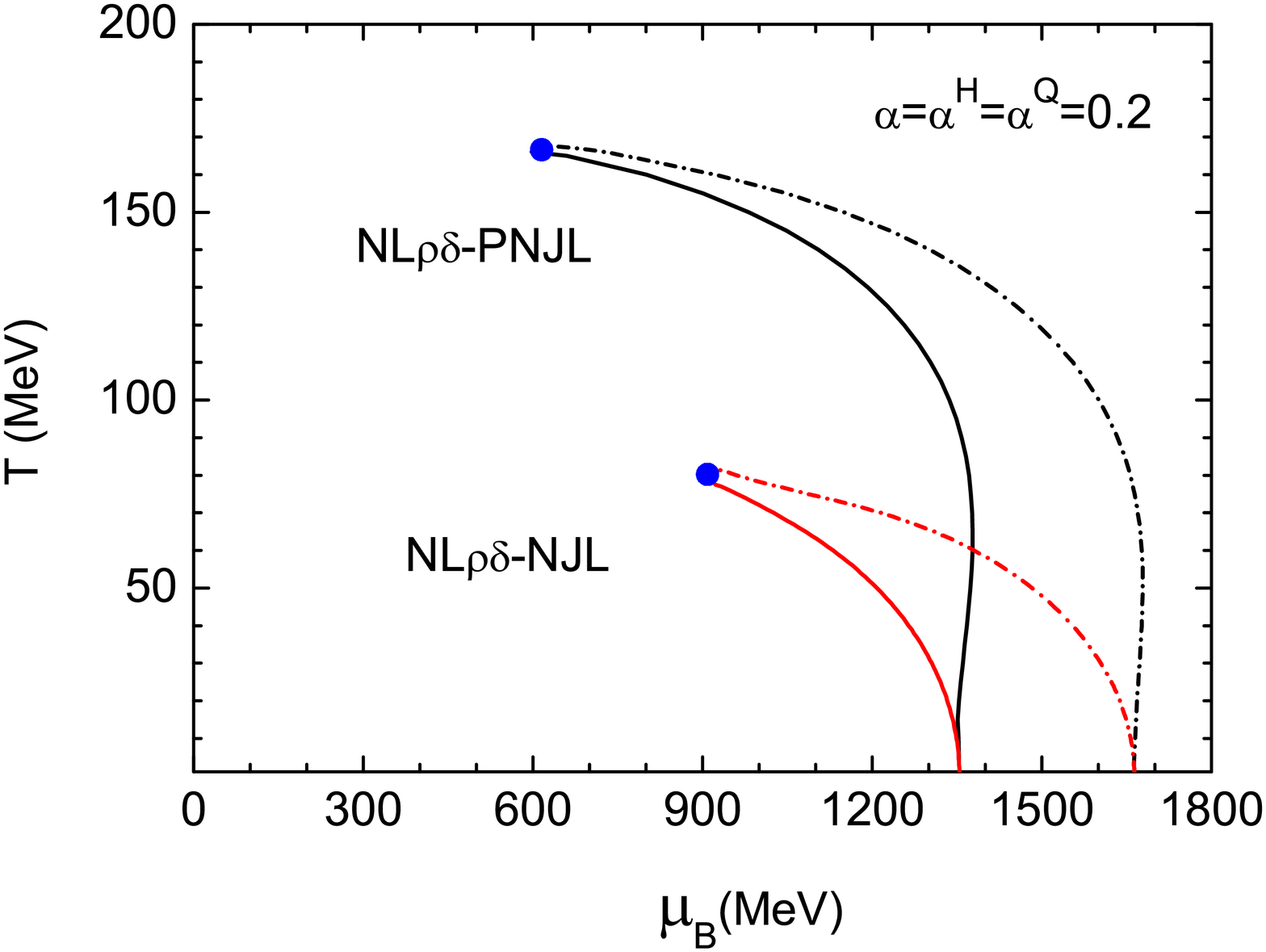}%{val1.ps}
}
\caption{\label{fig:T-Mu-with-delta-alpha=02} (Color on line) Phase
diagram in  $T-\mu_B^{}$ plane for asymmetry matter with the global
asymmetry parameter $\alpha=0.2$, \cite{Shao112}. Full lines $\chi=0$, dash-dotted lines 
$\chi=1$, see text.}
%\end{center}
\end{figure}

In Fig.s ~\ref{fig:T-Mu-with-delta-alpha=0} and
~\ref{fig:T-Mu-with-delta-alpha=02} we plot the phase transition
curves for symmetric and asymmetric matter with the Hadron-NJL and Hadron-PNJL models. 
In the hadron sector we use RMF interactions, see the note ~\cite{hadint}, already tested
in heavy ion collisions at lower energies.

At low
temperatures a clear earlier onset of the transition is observed for
isospin asymmetric matter $\alpha=0.2$ (see full lines of
Fig.~\ref{fig:T-Mu-with-delta-alpha=02}). In the $T-\rho_B^{}$
plane we see the onset of the transition moving from $6.5$ down to
$4.5$ times the saturation density,~\cite{Shao112}, as will be seen below in the
Fig. ~\ref{fig:T-RHO-PNJL-alpha=02} .

 For the NJL model
with only chiral dynamics, no physical solution exists when the
temperature is higher than
 $\sim80$
MeV.  The corresponding temperature is enhanced to about $\sim166$
MeV with the Hadron-PNJL model,
 which is closer to the phase transition (crossover)
temperature given by full lattice calculations at zero or small
chemical potential.
  We also note that the expected
$Critical-End-Point$ in the Hadron-PNJL scheme appears at a quark
chemical potential $\mu_q=\mu_B/3\sim200$MeV, just above the
critical temperature, which may be accessible in lattice
calculations.

Comparing Fig.s ~\ref{fig:T-Mu-with-delta-alpha=0} and
~\ref{fig:T-Mu-with-delta-alpha=02} 
we remark that in both NJL and PNJL
cases the region around the $Critical-End-Points$ is almost not
affected by isospin asymmetry contributions, which are relevant at
lower temperatures and larger chemical potentials.

\begin{figure}[htbp]
%\begin{center}
\resizebox{0.4\textwidth}{!}{%
  \includegraphics{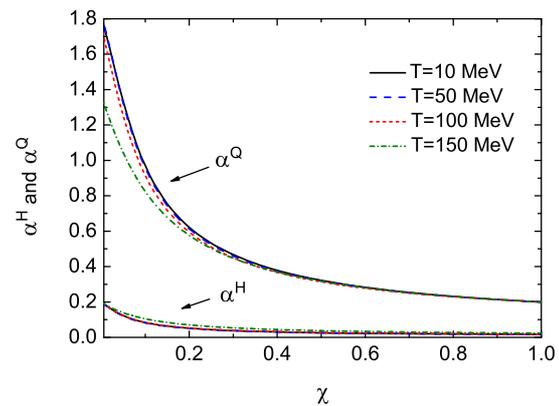}%{val1.ps}
}
\caption{\label{fig:kai-alphaQ-NLrho-with-delta-alpha=02} (Color on
line)
 The behavior of
local asymmetric parameters $\alpha^H$ and $\alpha^Q$ inside the mixed
phase for several values of temperature. Parameter set
$NL\rho\delta$ is used in the calculation \cite{Shao112}.}
%\end{center}
\end{figure}

\subsection{Isospin Distillation and Isospin Trapping} 

For symmetric matter there is only one phase-transition line in the
$T-\mu_B^{}$ plane, independent
of the quark fraction $\chi$. At variance, for asymmetric matter,
the phase transition curve varies for different quark fraction
$\chi$. The phase transition curves  in
Fig.~\ref{fig:T-Mu-with-delta-alpha=02} are obtained with $\chi=0$ (solid line)
and $1$ (dash-dotted line),  representing the beginning and the end of the hadron-quark
transition, respectively. The reason is that we have an important
Isospin Fractionation (Distillation) effect, i.e., an enhancement of
the isospin asymmetry in the quark component inside the mixed phase,
as reported in
Fig.~\ref{fig:kai-alphaQ-NLrho-with-delta-alpha=02}, where the
asymmetry parameters in the two components are plotted vs. the quark
fraction $\chi$. In asymmetric matter at a fixed temperature, along
the transition path, i.e. increasing the quark fraction, pressure
and chemical potential change and the two coexisting phases have
different asymmetry. The effect is particularly large at the
beginning of the mixed phase, i.e. for quark concentrations below
$50~\%$, which is in fact the region expected to be better probed in
heavy ion collisions at intermediate energies.

In the hadron phase the neutrons at high density are seeing a large repulsive symmetry term
and so are expected to be quickly emitted during the heavy ion collision. At variance, at the onset 
of the mixed phase the neutrons will be kept in the system since a more isospin asymmetric
quark phase should be formed. This is what we call "Isospin Trapping" and we expect
observable effects of that in the hadronization during the expansion step, see the final section.

We stress again that the isospin distillation effect is due to the large
difference in the symmetry terms in the two phases since all the
used quark effective models do not have explicit interaction
isovector fields.
We have checked that even with isovector terms in the
PNJL lagrangian, the effect is still there although a little reduced
\cite{Shao12}. 
In any case the strength of the couplings in the QCD
effective lagrangian is completely unknown and also for this reason
experiments at NICA appear very relevant.

\subsection{Vector quark interactions and existence of hybrid
neutron stars}

It appears natural to further investigate the role of vector
interactions in the quark effective models \cite{Shao12,Shao13}. We
remind that in nuclear matter the vector interactions lead to
fundamental properties, like the saturation point and the symmetry
energy in isospin asymmetric systems. In ref. \cite{Shao12} we have
seen how an isovector-vector term in the quark phase can affect the
isospin distillation mechanism.
 
We briefly discuss now some results
obtained when the isoscalar--vector interaction channel in the quark
sector is turned on in the (P)NJL models.
 With increasing the ratio $R_V=G_V/G$ of the vector/scalar coupling constants,
 due to the repulsive contribution of the
isoscalar--vector channel to the quark energy and, as a consequence,
to the chemical potential, the phase-transition curves are moving
towards higher values of density/chemical potential \cite{Shao12,Shao13}.

A larger
repulsion in the quark phase is essential for the existence of
massive hybrid neutron stars. 
However a limit appears to be the
impossibility of reaching the onset densities of the mixed phase in
the inner core for large values of the vector coupling.
A massive hybrid neutron star can
be supported in the range 0.1-0.3 of the $R_V$ ratio and good
agreement with recent data for the Mass-Radius relation is 
obtained \cite{Shao13}. In this more general respect the NICA data on properties of the mixed phase
in Heavy Ion Collisions are of great importance.

%%%%%%%%%%%%%%%%%%%%%%%%%%%%%%%%%%%%%%%%%%%%%%%%%%%%

\section{Conclusion: experimental proposals for the NICA program}

Asymmetric matter is interesting for two main reasons: i)
Earlier onset of the mixed phase; ii) Isospin distillation to the
quark phase. We have seen that noticeable effects can be observed
even for relatively low asymmetries, $\alpha \sim 0.2-0.3$, thus
experiments with stable heavy ions would be sufficient. Of course the
availability of very neutron-rich unstable beams in this energy
range could be very important.

The best region of the nuclear matter phase diagram to observe the
mixed phase appears to be at temperatures $T \sim 50-100 MeV$,
chemical potentials $\mu_B \sim 1200-1800 MeV$ (densities $\rho \sim
3-6\rho_0$). This can be achieved in HIC (fixed target) at beam
energies $3-10 AGeV$, $\sqrt{s_{NN}} \sim 3-5 GeV$, well inside the
reach of the NICA project.

\subsection{Suggested Observables}

As stressed before, since an equation of state able to describe the two phases is not presently available we cannot present results of a transport simulation of  heavy ion collisions with hadron-quark transitions. However from our knowledge of the hadronic collisions and from the results of the Two-EoS model discussed here we can suggest some possible interesting experiments in the NICA energy range. An important  general point is to concentrate on particles emitted at high transverse momentum. Indeed this kind of emission essentially occurs during the first stage of the collision, when a significant degree of compression is reached. Thus these particles are expected to keep track of the features of asymmetric high density matter. We can list:

\textbf{Onset of quark degrees of freedom:}

 i)In heavy ion collisions at intermediate energies, the high pressure reached in the high density phase induces an anisotropic  azimuthal distribution of the emitted particles, with a negative elliptic flow $v_2$.
The onset of the quark phase would cause an EoS softening, due to the lack of repulsive vector interactions in the quark matter joint  to an increase of the degrees of freedom. Then we would expect to observe a less negative $v_2$, with respect to theoretical predictions considering only hadronic matter. 

ii)Since hadrons may also be produced from the quark phase, one could observe, already at intermediate energies, the onset of $n_q$-scaling of the flows of the emitted particles, already seen at RHIC-LHC energies \cite{Fries08}. 
Of course one has to consider that lowering the beam energy the impact of the hadronic rescattering is expected to increase and one should argue that most of the flow is build-up in the early quark phase. Nonetheless the $n_q$ acquires a stronger significance at higher momenta ($>1$ GeV) that are associated to the early stage dynamics and are marginally affected by the later rescattering. A quantitative study is certainly necessary in this direction.

%\vskip 0.5cm

%
\begin{figure}[htbp]
%\begin{center}
\resizebox{0.4\textwidth}{!}{%
  \includegraphics{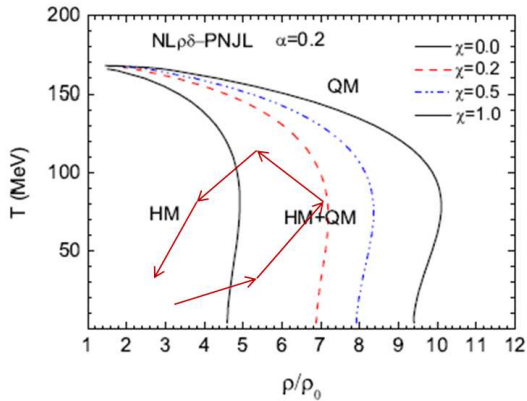}%
}
\caption{\label{fig:T-RHO-PNJL-alpha=02} (Color on line) Phase
diagram in $T-\rho_B^{}$ plane in the Two-EoS model for asymmetric
matter with the global asymmetry parameter $\alpha=0.2$. $\chi$
represents the fraction of quark matter.}
%\end{center}
\end{figure}

 \textbf{Isospin Distillation, $d$-rich quark phase.} 
At the NICA beam energies we can figure out a compression-expansion path inside the mixed phase like the one shown in the Fig.~\ref{fig:T-RHO-PNJL-alpha=02} (the arrow line, just pictorial representation).
We can expect a remaining influence on the final hadron formation, according to the following considerations:

i)The isotopic content of light cluster emission is seen to increase with the beam energy, owing to the higher compression reached and to the  more repulsive isovector interaction \cite{Xiao09}. However, if quark matter starts to co-exist with hadronic matter, the latter becomes more symmetric. Then one should observe a sudden inversion in the trend of  emission  of fast neutron-rich clusters with increasing beam energy, to be probed looking at the $n/p, ^3H/^3He...$ ratios at high transverse kinetic energy.

ii) Correspondingly, one should observe an enhanced production of isospin-rich nucleon resonances (and subsequent decays), originating from the quark phase.

iii) Related to the previous point, one would expect an anomalous increase of $\pi^-/\pi^+, K^0/K^+$ yield ratios for mesons coming from high density regions, i.e. with high transverse momentum $p_T$. 
We note that in absence of the phase transition, these ratios would keep smaller due to the fast neutron emission, as mentioned above, which inhibits the production of $d$-rich mesons in inelastic nucleon-nucleon collisions.

All these effects will be more relevant in the lower energy part of
the mixed phase, where the isospin distillation appears larger. A
Beam Energy Scan procedure would be appropriate for the suggested
isospin observables in the beam energy range $3-10 AGeV$. Recently a
similar beam energy search for the mixed phase in Heavy Ion
Collisions has been suggested for other observables, like transverse
mass, rapidity distribution and strangeness production
\cite{Kizka15}. We propose here an extension to isospin anomalies
using colliding ions with large charge asymmetries.

%\vskip 0.5cm

Finally the NICA results will be important for tuning the vector
interactions in the quark sector, relevant for neutron star models
and in general for the development of a unified effective field
theory of the two phases.

%%%%%%%%%%%%%%%%%%%%%%%%%%%%%%%%%%%%%%%%%%%%%%%%%%%%%%%%%%%%%%


\begin{thebibliography}{99}

%%%%%%%%%%%%%%%%%%%%%%%%%%%%%%%%%%%%%%%%%%%%%%%%%%%%%%%%%%%%%%%%%%%%%
%\bibitem{surNPA1988} E, Suraud, M. Pi, P. Schuck, Nucl. Phys. A {\bf 482}, 187c (1988).

\bibitem{baranPR} V. Baran, M. Colonna, V. Greco and M. Di Toro,
Phys. Rep. {\bf 410}, 335 (2005).

\bibitem{Toro06}M. Di Toro, A. Drago, T. Gaitanos, V. Greco, and A. Lavagno,
 Nucl. Phys. A  {\bf 775}, 102 (2006).
\bibitem{Toro09}M. Di Toro \emph{et al.}, Prog. Part. Nucl. Phys. {\bf 62}, 389 (2009).
%%%%%%%%%%%%%%%%%%%%%%%%%%  Massimo
\bibitem{Liu11} B. Liu, M. Di Toro, G. Y. Shao, V. Greco, C. W. Shen, and Z. H. Li,
Eur. Phys. J. A {\bf 47}, 104 (2011).
\bibitem{Yagi2005} K. Yagi, T. Hatsuda, Y. Miska, \emph{Quark Gluon Plasma} Ch.3, 
Cambridge Univ. Press (2005).
%\bibitem{Karsch01} F. Karsch, E. Laermann, and A. Peikert, Nucl. Phys.  {\bf B605} 579 (2001).
%\bibitem{Cheng06} M. Cheng \emph{et al.}, Phys. Rev. D {\bf 74}, 054507 (2006).
%\bibitem{YAoki09} Y. Aoki, \emph{et al.}, J. High Energy Phys. {\bf 06}, (2009) 088.
%\bibitem{Borsanyi10}S. Bors\'{a}nyi \emph{et al.},  J. High Energy Phys. {\bf 09}, (2010) 073.


%\bibitem{Fodor02}Z. Fodor and S.D. Katz, Phys. Lett. B {\bf 534}, 87 (2002); J. High Energy Phys. {\bf 03}, (2002) 014.
%\bibitem{Fodor03} Z. Fodor, S.D. Katz, and C. Schmidt, J. High Energy Phys. {\bf 03}, (2007) 121.
%\bibitem{Elia09}M. D'Elia and F. Sanfilippo, Phys. Rev. D {\bf 80}, 014502 (2009).
%\bibitem{Ejiri08}S. Ejiri, Phys. Rev. D {\bf 78}, 074507 (2008).
%\bibitem{Clark07}M. A. Clark and A. D. Kennedy, Phys. Rev. Lett. {\bf 98}, 051601 (2007).
\bibitem{Fukushima11} K. Fukushima and T. Hatsuda, Rep. Prog. Phys. {\bf 74}, 014001 (2011).
\bibitem{Chodos74} A. Chodos \emph{et al}, Phys. Rev. D {\bf 9}, 3471 (1974).
\bibitem{Nambu61}Y. Nambu and G. Jona-Lasinio, Phys. Rev. {\bf 112} (1961),
345; Phys. Rev. {\bf 124}, 246 (1961).
%\bibitem{Toublan03}D. Toublan and J. B. Kogut, Phys. Lett. B {\bf 564}, 212 (2003).
%\bibitem{Werth05}S. B. R\"{u}ster, V. Werth, M. Buballa, I. A. Shovkovy, and D. H. Rischke, Phys. Rev. D {\bf 72}, 034004 (2005).
%\bibitem{Abuki06}H. Abuki and T. Kunihiro, Nucl. Phys. {\bf A768},
%118 (2006).


%\bibitem{Volkov84}M. K. Volkov, Ann. Phys. (N.Y.) {\bf 157}, 282 (1984).
%\bibitem{Hatsuda84}T. Hatsuda and T. Kunihiro, Phys. Lett. B {\bf 145}, 7
%(1984).
%\bibitem{Klevansky92} S. P. Klevansky, Rev. Mod. Phys. {\bf 64}, 649 (1992).
%\bibitem{Hatsuda94} T. Hatsuda and T. Kunihiro, Phys. Rep. {\bf 247}, 221
%(1994).
%\bibitem{Alkofer96} R. Alkofer, H. Reinhardt, and H.Weigel, Phys. Rep.
%{\bf 265}, 239 (1996).

\bibitem{Buballa05}
M. Buballa, Phys. Rep. {\bf 407}, 205 (2005).
%\bibitem{Rehberg95}P. Rehberg, S. P. Klevansky, and J. H\"{u}fner, Phys. Rev.
%C {\bf 53}, 410 (1996).

%\bibitem{Shovkovy03}I. Shovkovy, and Mei Huang, Nucl. Phys. {\bf B564}, 205
%(2003).
%\bibitem{Huang03}M. Huang  and I. Shovkovy, Nucl. Phys.  {\bf  A729}, 835
%(2003).
%\bibitem{Alford08}M. Alford, A. Schmit, K. Rajagopal, and  T. Sch\"{a}fer,
%Rev. Mod. Phys. {\bf 80}, 1455 (2008).

\bibitem{Fukushima04}K. Fukushima, Phys. Lett. B {\bf  591}, 277 (2004).
\bibitem{Ratti06}C. Ratti, M.A. Thaler, W. Weise, Phys. Rev. D {\bf 73}, 014019 (2006).
%\bibitem{Costa10} P. Costa, M. C. Ruivo, C. A. de Sousa, and H. Hansen,
%Symmetry {\bf 2(3)}, 1338, (2010)

\bibitem{Schaefer10}B-J. Schaefer, M. Wagner, J. Wambach,
 Phys. Rev. D {\bf 81}, 074013 (2010).

\bibitem{Borsanyi14}
  S. Borsanyi, Z. Fodor, C. Hoelbling, S.D. Katz, S. Krieg and K.K. Szabo,
  %``Full result for the QCD equation of state with 2+1 flavors,''
  Phys.\ Lett.\ B {\bf 730}, 99 (2014)
%\bibitem{Bazavov14}
 % A. Bazavov {\it et al.} [HotQCD Collaboration],
  %``Equation of state in ( 2+1 )-flavor QCD,''
  %Phys.\ Rev.\ D {\bf 90}, 094503 (2014)
%\bibitem{Herbst11} T. K. Herbst, J. M. Pawlowski, B-J. Schaefer,
% Phys. Lett. B {\bf 696}, 58 (2011)


%\bibitem{Kashiwa08}K. Kashiwa, H. Kouno, M. Matsuzaki,  and M. Yahiro, Nucl.
% Phys.  {\bf B662}, 26 (2008).
%\bibitem{Abuki08}H. Abuki, R. Anglani, R. Gatto, G. Nardulli and M. Ruggieri,
%Phys. Rev. D {\bf 78}, 034034 (2008).
%\bibitem{Fu08}W. J. Fu, Z. Zhang, Y. X. Liu, Phys. Rev. D {\bf 77}, 014006
%(2008).


\bibitem{Glendenning92}N. K. Glendenning, Phys. Rev. D {\bf 46},  1274  (1992).
\bibitem{Glendenning98}N. K. Glendenning and J. Schaffner-Bielich, Phys. Rev.
Lett. {\bf 81}, 4564 (1998); Phys. Rev. C {\bf 60}, 025803 (1999).
\bibitem{Burgio02}G. F. Burgio, M. Baldo, P. K. Sahu, and H.-J. Schulze, Phys.
 Rev. C {\bf 66}, 025802 (2002).
%\bibitem{Maruyama07}T. Maruyama, S. Chiba, H-J. Schulze, and T. Tatsumi, Phys.
%Rev. D {\bf 76}, 123015 (2007).
%\bibitem{Yang08}F. Yang and H. Shen, Phys. Rev. C {\bf 77}, 025801 (2008).
\bibitem{Shao10}G. Y. Shao and Y. X. Liu, Phys. Rev. C {\bf 82}, 055801 (2010).
\bibitem{Shao110}G. Y. Shao, Phys. Lett. B {\bf 704}, 343 (2011).

%\bibitem{Xu10}J. Xu, L. W. Chen, C. M. Ko, and B. A. Li, Phys. Rev. C {\bf 81},
% 055803 (2010).

\bibitem{Dexheimer101}V. A. Dexheimer, S. Schramm, Phys. Rev. C {\bf 81}, 045201 (2010).
%\bibitem{Dexheimer102}V. A. Dexheimer, S. Schramm, Nucl. Phys. B  {\bf 199}, 319 (2010).


\bibitem{Muller97}H. M\"{u}ller, Nucl. Phys. A  {\bf 618}, 349 (1997).
%\bibitem{Toro06}M. Di Toro, A. Drago, T. Gaitanos, V. Greco, and A. Lavagno,
% Nucl. Phys. A  {\bf 775}, 102 (2006).
%\bibitem{Toro09}M. Di Toro \emph{et al.}, Prog. Part. Nucl. Phys. {\bf 62}, 389 (2009).
\bibitem{Torohq11} M. Di Toro \emph{et al.},
Phys. Rev. C {\bf 83}, 014911 (2011).

\bibitem{Cavagnoli10} R. Cavagnoli, C. Provid\^{e}ncia, and D. P. Menezes,
 Phys. Rev. C {\bf 83}, 045201 (2011).

\bibitem{Shao111}G. Y. Shao, M. Di Toro, B. Liu, M. Colonna, V. Greco, Y. X. Liu, and S. Plumari,
Phys. Rev. D {\bf 83}, 094033 (2011).

\bibitem{Shao112} G. Y. Shao, M. Di Toro, V. Greco, M. Colonna,  S. Plumari, B. Liu, and Y. X. Liu,
Phys. Rev. D {\bf 84}, 034028 (2011).

\bibitem{hadint} In the RMF scheme here we present results with three effective interactions: i) $NL$,
for symmetric matter, without isovector meson contributions; ii) $NL\rho$, with the 
inclusion of the isovector-vector $\rho$ channel, which leads to a potential contribution to the symmetry energy linearly increasing with density; iii) $NL\rho\delta$, including also the isovector-scalar $\delta$ field, with a symmetry term even more repulsive at high density. Details can be found in \cite{baranPR}.

%\bibitem{Pagliara10} G. Pagliara and J. Schaffner-Bielich, Phys. Rev.
%D {\bf 81}, 094024 (2010).

\bibitem{Shao12} G. Y. Shao, M. Colonna, M. Di Toro, B. Liu, F. Matera,
 Phys. Rev. D {\bf 85}, 114017 (2012).

\bibitem{Shao13} G. Y. Shao, M. Colonna, M. Di Toro, Y.X. Liu, B. Liu,
 Phys. Rev. D {\bf 87}, 096012 (2013).

\bibitem{Fries08}
  R.J. Fries, V. Greco and P. Sorensen,
  %``Coalescence Models For Hadron Formation From Quark Gluon Plasma,''
  Ann.\ Rev.\ Nucl.\ Part.\ Sci.\  {\bf 58}, 177 (2008).

\bibitem{Xiao09} Z. Xiao et al.,
 Phys. Rev. Lett. {\bf 102}, 062502 (2009).

\bibitem{Kizka15} V.A. Kizka, V.S. Trubnikov, K.A. Bugaev, D.R.
Oliinychenko, 
{\it  "A possible evidence of the hadron-quark-gluon mixed phase formation in nuclear collisions"}, 
 arXiv:1504.06483v1.

%\bibitem{Blaschke10} D. Blaschke, J. Berdermann, R. Lastowiecki, Prog. Theor. Phys. Suppl. {\bf 186}, 81 (2010).
%\bibitem{Ohnishi11} A. Ohnishi, H. Ueda, T. Z. Nakano, M. Ruggieri, and K. Sumiyoshi,
%arXiv:1102.3753v2 [nucl-th].

%\bibitem{Liu11}B. Liu, M. Di Toro, G. Y. Shao, V. Greco, C. W. Shen, and Z. H. Li, Eur. Phys. J. A {\bf 47}, 104 (2011).

%\bibitem{Toro2006}M. Di Toro, A. Drago, T. Gaitanos, V. Greco, and A. Lavagno, Nucl. Phys.
% {\bf A775}, 102 (2006).
%\bibitem{Toro09}M. Di Toro \emph{et al.}, Prog. Part. Nucl. Phys.
%{\bf 62}, 389 (2009).
%\bibitem{Toro11}M. Di Toro \emph{et al.}, Phys. Rev. C {\bf 83},
%014911 (2011).
%*************************************
%\bibitem{Kita02} M. Kitazawa, T. Koide, T. Kunihiro, and Y. Nemoto,
%Prog. Theor. Phys. {\bf 108}, 929 (2002).
%\bibitem{Sasa07} C. Sasaki, B. Friman, and K. Redlich, Phys. Rev. D {\bf 75}, 054026 (2007).
%\bibitem{Fuku08} K. Fukushima, Phys. Rev. D {\bf 77}, 114028 (2008).
%**************************************
%\bibitem{fock}
%This is not surprising since in the hadron sector, within
%the Relativistic Mean Field approach, the Fock terms also give
%a substantial fraction of the effective coupling constants,
%\cite{greco01,baran05}.

%\bibitem{greco01} V.Greco, M. Colonna, M.Di Toro, G.Fabbri, F.Matera,
%           Phys. Rev. {\bf C64}, 045203 (2001).
%\bibitem{baran05} V.Baran, M.Colonna, V.Greco, M.Di Toro,
%Phys. Rep. {\bf 410}, 335 (2005), Sect.6.1

%\bibitem{Robner07}S. R\"{o}{\ss}ner, C. Ratti, and W. Weise, Phys. Rev. D {\bf 75},
%034007 (2007).
%\bibitem{Fukugita90}M. Fukugita, M. Okawa, and A. Ukava, Nucl. Phys.  {\bf B337},
%181 (1990).
%\bibitem{Hansen07}H. Hansen, W. M. Alberico, A. Beraudo, A. Molinari,
%M. Nardi, and C. Ratti, Phys. Rev. D {\bf 75}, 065004 (2007).






%%%%%%%%%%%%%%%%%%%%%%%%%%%%%%%%%%%%%%%%%%%%%%%%%%%%%%%%%%%%%%
\end{thebibliography}
\end{document}